\documentclass[letterpaper,12pt]{JHEP3} 
\usepackage{bm}
\usepackage{amsmath}
\usepackage{yfonts}
\usepackage{epsfig,latexsym}

\newcommand{\wn}{\textswab{w}}
\newcommand{\<}{\langle}
\renewcommand{\>}{\rangle}

\newcommand{\q}{\bm{q}}
\newcommand{\x}{\bm{x}}
\newcommand{\ka}{\bm{k}}

\newcommand{\stru}{\rule[-.2in]{0in}{.2in}}

\title{Coupling constant dependence of the
 shear viscosity in ${\cal N}=4$ supersymmetric 
Yang-Mills theory}

\author{
Alex Buchel$ ^{\, a b}$\\
$ ^a$Perimeter Institute for Theoretical Physics\\
Waterloo, Ontario N2J 2W9, Canada\\
$ ^b$Department of Applied Mathematics, 
University of Western Ontario\\
London, Ontario N6A 5B7, Canada\\
Email: \email{abuchel@perimeterinstitute.ca}
}
\author{
 James T. Liu\\
Michigan Center for Theoretical Physics,
Randall Laboratory of Physics\\
 The University of Michigan\\
Ann Arbor, MI 48109-1120, USA\\  
Email: \email{jimliu@umich.edu }
}
\author{Andrei O.~Starinets\\
Institute for Nuclear Theory, 
 University of Washington\\
Seattle, WA 98195, USA\\
Email: \email{starina@phys.washington.edu}
}

\preprint{ MCTP-04-38\\ INT-PUB 04-17 \\ hep-th/0406264 }

\abstract{Gauge theory - gravity  duality predicts that
the shear viscosity of  ${\cal N}=4$ supersymmetric 
$SU(N_c)$ Yang-Mills plasma at temperature $T$ in the limit of 
large $N_c$ and large  't Hooft coupling 
$g^2_{YM}N_c$ is independent of the coupling 
and equals to $\pi N_c^2 T^3/8$. 
In this paper, we compute the leading correction to the shear viscosity
in inverse powers of  't Hooft coupling using the $\alpha'$-corrected 
low-energy effective action of type IIB string theory.
We also find the correction to the ratio 
of shear viscosity to the volume entropy density (equal to 
$1/4\pi$ in the limit of infinite coupling). The correction to 
$1/4\pi$ scales as  $(g^2_{YM}N_c)^{-3/2}$ with a positive
coefficient.
}


\begin{document}

\section{Introduction}
\label{section_introduction}

Transport coefficients such as viscosity, diffusion constants, thermal 
and electric conductivities are the key ingredients in describing
the hydrodynamic regime of any medium \cite{landau}.
These coefficients are usually obtained from experiment rather than
computed from first principles of an underlying microscopic theory,
because the study of realistic strongly interacting systems often 
remains beyond the reach of currently available theoretical methods.

For finite-temperature quantum field theories, 
and thermal gauge theories in particular,  
computations based on the Boltzmann equation
in the regime of weak coupling $g\ll 1$ show 
\cite{Jeon:if, Jeon:1995zm, Arnold:2000dr, Wang:2002nb} 
that the shear viscosity
behaves as 
\begin{equation}
\eta \sim \frac{ C\, N^2_c\, T^3 }{ g^4 \log{1/g^2}}\,,
\label{visc_perturbative}
\end{equation}
where $N^2_c$ is the number of colors and 
$C$ is a large numerical coefficient.

Since the entropy scales as $S \sim N^2_c\, T^3\, V_3$, the ratio of 
shear viscosity to the volume entropy density $s=S/V_3$ behaves as
\begin{equation}
{\eta \over s} \sim  { 1 \over g^4 \log{1/g^2}} \gg 1 
\label{ratio_perturbative}
\end{equation}
in the regime of weak coupling.
On the other hand, hydrodynamic models used to describe 
elliptic flows observed in recent heavy ion 
collision experiments at RHIC seem to favor
small values of the ratio $\eta/s$  \cite{Teaney:2003pb, 
Shuryak:2003xe, Molnar:2001ux}.
This is not a contradiction, however,
since the experimental data were obtained 
for the range of energies where the coupling 
remains relatively large, whereas the result (\ref{ratio_perturbative})
is valid for small coupling. 

It is therefore desirable to obtain results for the viscosity-entropy 
ratio in the regime of intermediate and strong coupling.
Lattice simulations cannot address the issue of real-time dynamics 
directly facing (among other things) 
a formidable problem of analytic continuation.
(For indirect approaches, see \cite{Karsch:1986cq}, \cite{Aarts:2002cc},
\cite{Gupta:2003zh},
\cite{Nakamura:2004sy}.)

In the absence of more conventional methods, the AdS/CFT
(or gauge theory/gravity) duality conjecture 
in string theory \cite{Maldacena:1997re, Gubser:1998bc, Witten:1998qj, Aharony:1999ti}
emerged as a source of insights into the 
non-perturbative regime of thermal gauge theories.
The best studied example of the duality relates a four-dimensional 
finite-temperature 
 ${\cal N}=4$ 
$SU(N_c)$ supersymmetric Yang-Mills (SYM) theory
in the limit of 
large $N_c$ and large 't Hooft coupling $g^2_{YM}N_c$ to
the supergravity background corresponding to a stack of 
$N_c$  near-extremal black three-branes.

Using the  AdS/CFT conjecture, one is able to predict the 
behavior of the entropy of  ${\cal N}=4$  SYM 
in the regime of strong coupling 
\cite{Gubser:1996de}, \cite{Gubser:1998nz}. 
In the large $N_c$ limit, the entropy
is given by
\begin{equation}
S = {2\pi^2\over 3} N^2_c V_3 T^3 f(g^2_{YM} N_c)\,,
\label{entropy_strong}
\end{equation}
where the function $f(g^2_{YM} N_c)$
interpolates (presumably smoothly) 
between $1$ at zero coupling and $3/4$ at infinite coupling.
The strong coupling expansion for $f$ was obtained by 
Gubser, Klebanov and Tseytlin\footnote{Normalization convention for $g_{YM}$
adopted in the current version of this paper differs from the one used in 
\cite{Gubser:1998nz}. See Eq.~(\ref{alpha}) and footnote \ref{ftn}.} 
 \cite{Gubser:1998nz}
\begin{equation}
f(g^2_{YM} N_c) = {3\over 4} + 
{45\over 32} \zeta (3) (g^2_{YM} N_c)^{-3/2}+\cdots\,.
\label{f_strong}
\end{equation}
At weak coupling, $f$ can be found by using 
finite-temperature perturbation theory \cite{Fotopoulos:1998es}
\begin{equation}
f(g^2_{YM} N_c) = 1 - {3\over 2 \pi^2}  g^2_{YM} N_c +
{3+\sqrt{2}\over \pi^3}   ( g^2_{YM} N_c)^{3/2} +\cdots
\label{f_weak}
\end{equation}
Our goal in the present paper is to obtain the analogue of the 
strong coupling expansion (\ref{f_strong}) for the shear 
viscosity and for the ratio $\eta/s$.
In the limit of infinite $N_c$ and infinite 't Hooft coupling, the shear 
viscosity  of ${\cal N}=4$  SYM plasma was found to be \cite{Policastro:2001yc}
\begin{equation}
\eta =  {\pi\over 8} N^2_c T^3\,.
\label{viscosity_strong}
\end{equation}
Taking into account the leading term in (\ref{f_strong}), one may 
notice that \cite{Kovtun:2003wp}
\begin{equation}
{\eta\over s}  =  {1\over 4 \pi} + O \left( 1/N^2_c\right) + 
O \left( (g_{YM}^2 N_c)^{-3/2}\right)\,.
\label{ratio_strong}
\end{equation}
The shear viscosity (\ref{viscosity_strong}) 
was obtained by a number of interconnected methods
\cite{Policastro:2001yc}, \cite{Policastro:2002se}, \cite{Policastro:2002tn}, 
\cite{Kovtun:2003wp}
all of which relied on the validity of the 
conjectured gauge/gravity correspondence. In that sense, the result 
(\ref{viscosity_strong}) 
does not enjoy the status of a firmly established 
field-theoretical calculation and 
should be viewed as a prediction of a 
gauge/gravity duality for a particular supersymmetric theory\footnote{Over
the years, AdS/CFT has survived numerous checks at zero temperature
\cite{Aharony:1999ti}. 
For a check at finite temperature, see \cite{Policastro:2002tn}.}.

It was further found in \cite{Kovtun:2003wp} that the ratio
$\eta/s$ remains equal to $1/4\pi$ for 
other gauge/gravity duals
where $\eta$ and $s$ alone behave very 
differently compared to the basic example of  ${\cal N}=4$  SYM.
The same feature holds for theories dual to $M$-branes 
\cite{Herzog:2002fn, Herzog:2003ke}.
The calculations in \cite{Kovtun:2003wp} were done independently by using 
the traditional AdS/CFT strategy, and also by deriving a generic formula 
for $\eta/s$ by a method 
reminiscent of the old ``membrane paradigm'' 
approach\footnote{From the AdS/CFT point of view, this 
method presumably computes the lowest (hydrodynamic) quasinormal 
frequency of a given gravitational background. 
The corresponding link is not firmly established at the moment, see \cite{Son:2002sd}, 
\cite{Nunez:2003eq}.}.
It was later shown in  \cite{Buchel:2003tz} that the type II supergravity 
equations of motion guarantee that the generic value 
for $\eta/s$ is universally equal to  $1/4\pi$.
Whenever the near-horizon supergravity geometry dual to a finite temperature 
gauge theory allows for an extension to asymptotically flat 
space-time\footnote{Such an extension is not  known for explicit 
examples discussed in \cite{Buchel:2003tz}.}, 
the    universality of $\eta/s$ can be related \cite{Kovtun:2004de} 
to the universality  of low energy absorption cross sections for black holes
observed in \cite{Das:1996we}.

If the conjectured universality is indeed true, it is most 
probably associated with the regime of strong coupling 
in various theories. For weakly coupled theories, the ratio
 $\eta/s$ is typically very large, even at large $N_c$.
It was conjectured in \cite{Kovtun:2003wp} that  $1/4\pi$ is the lowest
possible value for  $\eta/s$ for all substances in nature. 
In view of this conjecture, finding the first nontrivial 
term in the strong coupling expansion for  $\eta/s$
is especially interesting.

Our paper is organized as follows. In Section  \ref{section_shear} 
we outline the method of computing the strong coupling expansion 
for shear viscosity of ${\cal N}=4$  SYM from supergravity.
In Section \ref{section_perturbation} we consider gravitational 
perturbations of the  $\alpha'$- corrected black brane metric
and compute 
the boundary term of the corresponding on-shell action.
In Section \ref{section_correction} we use the Minkowski AdS/CFT
prescription to find the retarded thermal correlator of a 
stress-energy tensor in the hydrodynamic approximation and compute 
the correction to the shear viscosity.
Finally, in  appendix \ref{section_appendix} we collect 
the coefficients of the effective action which are  
too cumbersome to appear in the main text.

\section{Shear viscosity of ${\cal N}=4$ SYM from supergravity}
\label{section_shear}

Shear viscosity is the transport coefficient that appears in the 
hydrodynamic constitutive relation for the spatial components 
of the stress-energy tensor (see \cite{Policastro:2002se}, 
\cite{Kovtun:2003vj} 
for more hydrodynamics preliminaries).

Among the methods developed  in 
\cite{Policastro:2001yc}, \cite{Policastro:2002se},
 \cite{Kovtun:2003wp} for computing the 
shear viscosity from supergravity, the most straightforward 
and the least 
technically complicated one \cite{Policastro:2001yc} 
is based on the Kubo relation
\begin{equation}
\eta = \lim_{\omega \rightarrow 0} {1\over 2 \omega} \int d t \,  d^3\, x\,
e^{i \omega t} \langle [ T_{xy}(x) T_{xy}(0)]\rangle \,.
\label{kubo}
\end{equation}
The Kubo formula is a particular
 case of the celebrated fluctuation-dissipation 
theorem which relates fluctuations of a  
medium  to the response of the medium to an external action.
The formula expresses transport coefficients 
of a  slightly non-equilibrium 
system in terms of the real-time
 correlation functions computed in an equilibrium thermal ensemble.
Eq.~(\ref{kubo}) can be written in the form
 \begin{equation}
\eta = \lim_{\omega \rightarrow 0} {1\over 2 \omega i} \Biggl[ 
G_{xy,xy}^A (\omega,0) - G_{xy,xy}^R (\omega,0)\Biggr]\,,
\label{RA}
\end{equation}
where the retarded Green's function is defined as
\begin{equation}
  G_{\mu\nu,\lambda\rho}^R (\omega, \q)
  = -i\!\int\!d^4x\,e^{-iq\cdot x}\,
  \theta(t) \< [T_{\mu\nu}(x),\, T_{\lambda\rho}(0)] \>\,,
\label{retarded}
\end{equation}
and $G^A(\omega,\q) = (G^R (\omega, \q))^*$.

To compute the hydrodynamic limit 
of the retarded correlation function (\ref{retarded})
from supergravity, one can use either the Minkowski AdS/CFT prescription 
\cite{Son:2002sd}, \cite{Herzog:2002pc},  
 or the method based on calculating the absorption 
cross-section of gravitons by non-extremal three-branes 
\cite{Policastro:2001yb, Policastro:2001yc}.
In the absence of $\alpha '$ corrections, to
the lowest order in $\omega/T \ll 1$, $q/T \ll 1$ 
the retarded correlation function is given by  \cite{Policastro:2002se}
 \begin{equation}
 G_{xy,xy}^R(\omega,q) = - {N^2_c T^2\over 16} 
( i 2 \pi T \omega + q^2 +\cdots) \,.
\end{equation}
From   Eq.~(\ref{RA}) one finds the shear viscosity 
 \begin{equation}
\eta = {\pi\over 8} N^2_c T^3 + O \left( 1/N^2_c\right) + 
O \left( (g_{YM}^2 N_c)^{-3/2}\right)\,.
\end{equation}
Corrections to the shear viscosity in powers of the inverse 't Hooft
coupling correspond on the string theory side of AdS/CFT duality to 
$\alpha '$  corrections to classical general relativity. 
The precise relation between the expansion parameters in supergravity and 
Yang-Mills theory is given by\footnote{\label{ftn} The normalization convention 
$L^4/\alpha'\,^2=4\pi g_s N_c = g_{YM}^2 N_c$ \cite{Aharony:1999ti} 
used in the current version of this paper differs
from the normalization $L^4/\alpha'\,^2 = 2 g_{YM}^2 N_c$ used in the earlier
versions and in some other works by a factor of 2. In general, this normalization depends
on the normalization of the generators of the gauge group: $g^2_{YM} = 2\pi g_s/c$
 for $\mbox{tr}\, (T^a T^b) = c\, \delta^{ab}$. See e.g. Section 13.4 of \cite{kiritsis}.
 } 
\cite{Aharony:1999ti}
 \begin{equation}
{\alpha '\over L^2} = {1\over \sqrt{g^2_{YM} N_c}}\,.
\label{alpha}
\end{equation}
Since the leading order stringy correction to type IIB supergravity 
 is proportional to $\alpha '\,^3$, on the gauge theory side 
we expect the coupling constant 
 correction to shear viscosity to scale as $(g^2_{YM} N_c)^{-3/2}$.
At the same time, 
finding  $1/N^2_c$ corrections requires quantum gravity calculations 
in the  black three-brane background.

In this paper, we use the prescription of 
\cite{Son:2002sd} to compute the retarded correlator  $G_{xy,xy}^R$
from the $\alpha '$-corrected black three-brane metric found in 
\cite{Gubser:1998nz}. The coupling constant correction to 
viscosity then follows from the Kubo formula (\ref{RA}).

\section{Perturbations of the  $\alpha '$-corrected 
 near-extremal three-brane background}
\label{section_perturbation}

We start with the tree level type IIB low-energy effective action in ten
dimensions taking into account the leading order string corrections 
\cite{Grisaru:vi,Gross:1986iv}
\begin{equation}
S =
 {1\over 2 \kappa_{10}^2 } \int d^{10} x \sqrt g
\ \bigg[ R - {1\over 2} (\partial \phi)^2 - {1 \over 4 \cdot 5!} 
  (F_5)^2  +\cdots+ 
\  \gamma \ e^{- {3\over 2} \phi}  W + \cdots\bigg]   \  ,
\label{action}
\end{equation}
 \begin{equation}  
  \gamma= { 1\over 8} \zeta(3)(\alpha')^3 \ , 
 \label{gamma}
\end{equation}
where $W$ may be chosen in an appropriate scheme to be proportional to
the fourth power of the Weyl tensor  \cite{Gubser:1998nz}
\begin{equation}
W =  C^{hmnk} C_{pmnq} C_{h}^{\ rsp} C^{q}_{\ rsk} 
 + {1\over 2}  C^{hkmn} C_{pqmn} C_h^{\ rsp} C^{q}_{\ rsk}\  . 
\label{w}
\end{equation}
It is helpful to rewrite $W$ in the form \cite{deHaro:2003zd}
\begin{equation}
W = B_{i j k l} \left( 2 B^{i k l j} - B^{l i j k} \right), \ \ \ \ \ \
B_{i j k l} = C^m{}_{i j n} \, C^n{}_{l k m}.
\end{equation}
In (\ref{action}), ellipses stand for other fields not essential 
for the present analysis.
As usual, 
it is assumed that  the self-duality condition on $F_5$
 is imposed after the equations of motion are derived.
The form of the action (\ref{action}) 
and subtleties associated with the self-duality 
condition on the five-form are discussed in \cite{Gubser:1998nz}, 
\cite{deHaro:2003zd}.
We emphasize that the possibility of additional corrections
of order $(\alpha')^3$ associated with the five-form is not excluded.
However, indirect arguments presented in \cite{Gubser:1998nz}, 
\cite{deHaro:2003zd}
suggest that these corrections, if present, will not affect the near-horizon
geometry of the black three-brane solution.
We proceed with our calculation while ignoring this potential problem.

\subsection{The  $\alpha'$-corrected black three-brane background 
and its $S^5$ reduction}
\label{reduction}

In the absence of  $\alpha'$ corrections, the metric 
describing the non-extremal black three-brane solution
trivially factorizes into the five-sphere part and the ``AdS'' part.
However, this is no longer the case for the  $\alpha'$- corrected solutions,
as was first emphasized by Pawelczyk and Theisen \cite{Pawelczyk:1998pb}.
Thus in considering both the  $\alpha'$- corrected black brane solution 
and its fluctuations, the correct procedure is to work in ten dimensions 
and then perform the $S^5$ reduction to obtain the five-dimensional 
asymptotically AdS geometry. 
An additional subtlety comes from dealing with the self-dual five-form
$F_5$, which identically squares to zero by virtue of ten-dimensional
self-duality.

The ten-dimensional metric can be taken in the form of a 
standard ansatz used for the $S^5$ reduction of type IIB supergravity
(see e.g. \cite{Bremer:1998zp})
\begin{equation}
  ds^2_{10}  = e^{-{ 10 \over 3}\nu (x)} g_{5mn} (x) dx^m dx^n
 + e^{2 \nu(x)} d\Omega_5^2\,, 
\label{10m}
\end{equation}
where the five-dimensional asymptotically AdS metric  $g_{5mn} (x)$
has a general  form consistent with the symmetries of the problem\footnote{In
 a dual thermal field theory, 
Poincare invariance is broken, but the system is assumed to be isotropic.}  
\begin{equation}
ds_5^2={r_0^2\over u} e^{c(u)} \left( - f e^{a(u)} dt^2 + 
 dx^2 + d y^2 + d z^2 \right)  + {du^2\over 4 u^2 f} e^{b(u)}\,.
\label{mu5u}
\end{equation}
Here $f(u)=1-u^2$, $r_0$ is the parameter of non-extremality of the 
black brane geometry, and we set the ``AdS radius''
 $L$ to one. 
When $\gamma =0$ in (\ref{action}),
the metric (\ref{10m}) with $\nu=0$, $a=0$, $b=0$, $c=0$  
is the standard solution of type IIB low-energy equations of motion 
describing the near-horizon limit 
of non-extremal three-branes.  Corrections to that
solution can be found by solving equations of motion perturbatively 
in $\gamma$. To leading order in  $\gamma$, functions 
 $a$, $b$, $c$, $\nu$  were found in \cite{Gubser:1998nz}, 
\cite{Pawelczyk:1998pb}
\begin{subequations}
\begin{eqnarray}
a(u) &=& -15\, \gamma\, (5 u^2+5 u^4-3u^6)\,,\stru \\
b(u) &=& 15\, \gamma\, (5 u^2+5 u^4-19 u^6)\,, \stru \\
c(u) &=& 0\,, \stru \\
\nu (u) &=& {15 \gamma\over 32} u^4 (1+u^2)\,.
\end{eqnarray}
\end{subequations}
The Hawking temperature corresponding to the metric (\ref{10m}) is 
\begin{equation}
T = {r_0\over \pi} \left( 1 + 15 \gamma \right)\,.
\label{hawking}
\end{equation}
The dilaton $\phi$ also receives  $\alpha'$ corrections \cite{Gubser:1998nz}. 
Since to leading order in $\gamma$ they do not mix with the gravitational 
perturbations which we are considering in this paper, we do not write 
them explicitly.

Substituting the ansatz  (\ref{10m}) into the action 
 (\ref{action}) we find
\begin{equation}
S_5 = {\pi^3 \over 2 \kappa_{10}^2}
\int d^5 x \sqrt{g_5} \bigg[ R_5 + 
20 e^{-{ 16 \over 3}\nu}  - { 1 \over 2} (\partial \phi)^2
- { 40 \over 3}(\partial \nu)^2 -  8 e^{-{ 40 \over 3}\nu}
 + \gamma    e^{ - {3\over 2} \phi - {10\over 3} \nu }\,   W (u, \nu, \gamma)
\bigg]\ ,
\label{acti}
\end{equation}
where the ten-dimensional
gravitational constant $\kappa_{10} = \sqrt{8 \pi G}$ is related to the
number $N_c$ of coincident branes and the AdS radius $L$ by
 $\kappa_{10} = 2 \pi^2 \sqrt{\pi} L^4/N_c$.

For small $\nu \sim \gamma$ we have
\begin{equation}
 e^{-{10\over 3}\nu} W (u, \nu, \gamma) = 180 u^8 
 +   1800 u^8 \nu (u)  + O (\nu^2)\,.
\end{equation}
(Note that the expression for  $W (u, \nu, \gamma)$ is not merely the 
five-dimensional $W$ computed with the metric  (\ref{mu5u}).)

As mentioned above, there is a subtlety in reducing the term involving the 
five-form field strength. The correct procedure is to assume that $F_5$ has
components only along the five-sphere, and double that contribution in the 
effective ten-dimensional action. Another way of saying this
is that the term $-8 e^{-{ 40 \over 3}\nu}$ was added to 
(\ref{acti}) by hand to ensure that, upon variation,
the correct reduced equations of motion 
in five dimensions are reproduced.

\subsection{Metric perturbations}

The coupling between the boundary value of the 
graviton and the stress-energy tensor of a gauge theory
is given by $h^x_y (x) T_x^y/2$.
According to the AdS/CFT prescription, 
in order to compute the retarded thermal 
two-point function of the components of the 
stress-energy tensor entering the Kubo formula  (\ref{RA}),
we should add a small bulk perturbation $h_{xy}(u,x)$ 
to the metric (\ref{10m}),
and compute the on-shell action as a functional of its boundary value 
$h_{xy}(x)$.

Simple symmetry arguments 
\cite{Policastro:2002se} show that for a perturbation of this type and 
a metric of the form (\ref{10m}) all other components of a 
generic perturbation  $h_{\mu\nu}$ can be consistently set to zero.
(We have checked explicitly that to linear order in the perturbation, the 
equation for $h_{xy}$ decouples from all other equations describing a generic
perturbation of the background (\ref{10m}).)

It will be convenient\footnote{In the absence of $\alpha'$-corrections 
the effective action of $\varphi (u,x)$ is that of a
minimally coupled scalar in the AdS black hole background.} to
introduce a field  $\varphi (u,x)$
\begin{equation}
\varphi (u,x) = {u\over r_0^2} \, h_{xy} (u,x)
\end{equation}
(note that $h^x_y = g^{xx} h_{xy} = e^{{10 \nu\over 3}} \varphi (u,x) =
 \varphi (u,x) + O(u^4)$)
and use the Fourier decomposition
\begin{equation}
 \varphi (u,x) = \int\! {d^4 k\over (2\pi )^4} e^{-i\omega t + i \ka \cdot\x}
\varphi_k (u)\,.
\label{fourier}
\end{equation}
Since in   (\ref{RA})  we need the correlator at 
vanishing spatial momentum,
 it will be sufficient to consider 
perturbations which depend on the radial 
coordinate and time only.

\subsection{The effective action}

Substituting the perturbed $\alpha'$-corrected black brane metric 
into the ten-dimensional action  (\ref{action}) and performing the 
five-sphere reduction keeping in mind the subtleties mentioned in
Section \ref{reduction}, we obtain the effective action for the field 
$\varphi_k (u)$. To quadratic order in $\varphi_k (u)$ and linear order in 
$\gamma$ this action is given by 
\begin{equation}
S_5 = S^{(0)} + S^{(2)}\,, 
\end{equation}
where $ S^{(0)}$ is independent of $\varphi_k (u)$, and 
\begin{eqnarray}
S^{(2)} &=& { N^2_c \over 8 \pi^2 } \int  {d^4 k \over (2 \pi)^4}
 \int_0^1 du \Biggl[
    \, A\,  \varphi_{k}''\varphi_{-k} +
 B\,  \varphi_{k}'\varphi_{-k}'  + 
 C \, \varphi_{k}'\varphi_{-k} 
 \nonumber \\
 &+& D\, \varphi_{k}\varphi_{-k} 
  + E\, \varphi_{k}''\varphi_{-k}'' 
+ F\,  \varphi_{k}''\varphi_{-k}'\Biggr]\,. 
\label{s2}
\end{eqnarray}
The coefficients $A, B, C, D, E, F$ are even functions of the momentum.
They are given 
explicitly in appendix \ref{section_appendix}.

Variation of $S^{(2)}$ leads to 
\begin{equation}
\delta S^{(2)} = {N^2_c \over 8\pi^2} \int  {d^4 k \over (2 \pi)^4}
\;  \Biggl[\;\;\;  \int_0^1 du \left[ EOM \right] \delta \varphi_{-k}
+ \left( {\cal B}_1  \delta  \varphi_{-k} +
 {\cal B}_2  \delta  \varphi_{-k}'\right) \Biggl|_{0}^{1} \;\;\;   \Biggr]\,, 
\label{varact}
\end{equation}
where the coefficients of the boundary term are given by
\begin{equation}
 {\cal B}_1 = 
 - (A  \varphi_{k})' + 2 B  \varphi_{k}' + C  \varphi_{k} -
2 (E  \varphi_{k}'')' + F \varphi_{k}'' - (F  \varphi_{k}')'\,,
\end{equation}
\begin{equation}
 {\cal B}_2 =  A  \varphi_{k} + F  \varphi_{k}' 
+ 2 E \varphi_{k}''\,,
\label{boundary2}
\end{equation}
and EOM denotes the left hand side of the 
Euler-Lagrange equation
\begin{equation}
A \varphi_{k}'' + C \varphi_{k}' + 2 D \varphi_{k} - {d\over d u}
\left( 2 B \varphi_{k}'  + C \varphi_{k} +  F \varphi_{k}''\right)
 +  {d^2\over d u^2}
\left( A \varphi_{k}  + 2 E \varphi_{k}'' +  F \varphi_{k}'\right) = 0 \,.
\label{ele}
\end{equation}
In order to have a well-defined variational principle, one 
has to add a generalized Gibbons-Hawking boundary 
term to the action  (\ref{s2}). 
Variation of this additional term 
is supposed to cancel  the contribution 
${\cal B}_2  \delta  \varphi_{-k}'$ in  (\ref{varact}) 
without leading to
any new boundary terms involving first and higher 
derivatives of  $\varphi_{k}$.
In the absence of higher-derivative terms in the action, this is achieved by 
adding the standard  Gibbons-Hawking term \cite{Gibbons:1976ue}
proportional to the trace of the extrinsic curvature of the boundary,
i.e. by adding the boundary term $-A\varphi_{k}\varphi_{-k}'$ to the action
(\ref{s2}). 
Similarly, one can cancel the second term in  
${\cal B}_2$ by adding the boundary term 
 $-F\varphi_{k}'\varphi_{-k}'/2$. However, the last term in 
(\ref{boundary2})
poses a problem: adding a term  $-2E\varphi_{k}''\varphi_{-k}'$
to cancel it would upon variation produce second
derivatives of  $\delta  \varphi_{-k}$. This problem
is not unexpected because we have a theory whose equation of motion is 
a fourth order differential equation. 
In our case, however, the difficulty can be avoided
since we intend to treat the higher-derivative term as a perturbation, 
and we only solve the equation of motion (\ref{ele}) perturbatively 
to the first non-trivial 
order in $\gamma$. This means that effectively we are still 
solving a second order equation.
Let us write  Eq.~(\ref{ele}) in the form 
\begin{equation}
\varphi_{k}'' + p_1 \varphi_{k}' + p_0 \varphi_{k} = O(\gamma)\,, 
\end{equation}
where all $\gamma$-dependent terms are exiled to the right.
Now, if we add to the action (\ref{s2}) the generalized  Gibbons-Hawking term
\begin{equation}
{\cal K} = -A\varphi_{k}\varphi_{-k}'-{F\over 2} 
\varphi_{k}'\varphi_{-k}' + E\left(p_1 \varphi_{k}'
  + 2 p_0 \varphi_{k}\right)\varphi_{-k}'\,,
\label{gibbons}
\end{equation}
and note that the coefficient $E$ is proportional to $\gamma$,
then upon variation we find that the boundary term involving 
 $\delta  \varphi_{-k}'$ vanishes on shell up to  terms of 
order $\gamma^2$. Thus, to 
linear order in $\gamma$, the variational problem is now
well-posed.

The bulk action (\ref{s2}) can be rewritten in the form 
\begin{equation}
S^{(2)} = {N^2_c\over 8\pi^2} \int  {d^4 k \over (2 \pi)^4}
 \int_0^1 du \left( \partial_u {\cal B} + {1\over 2}
 \left[ EOM \right] \varphi_{-k} \right)\,,
\end{equation}
where
\begin{eqnarray}
{\cal B} &=& -{A'\over 2}  \varphi_{k} \varphi_{-k}  + 
 B  \varphi_{k}' \varphi_{-k}  +
{C\over 2}   \varphi_{k} \varphi_{-k}
-E'  \varphi_{k}'' \varphi_{-k}
+E  \varphi_{k}'' \varphi_{-k}'
- E  \varphi_{k}''' \varphi_{-k}\nonumber \\
&+& {F\over 2}   \varphi_{k}' \varphi_{-k}'
- {F'\over 2}  \varphi_{k}' \varphi_{-k}
\end{eqnarray}
and $EOM$ denotes the left hand side of (\ref{ele}).
Adding to this the generalized Gibbons-Hawking term  (\ref{gibbons}),
we observe that the complete on shell action reduces to a boundary term
\begin{equation}
 \int  {d^4 k \over (2 \pi)^4} \;  {\cal F}_k 
\Biggl|_{0}^{1}\,,
\end{equation}
where
\begin{eqnarray}
 {\cal F}_k &=& {N^2_c r_0^4\over 8 \pi^2} \Biggl[ \left( B - A\right) \varphi_{k}' \varphi_{-k} + {1\over 2}
 \left( C - A'\right) \varphi_{k} \varphi_{-k} -E'  \varphi_{k}'' \varphi_{-k}
+E  \varphi_{k}'' \varphi_{-k}' \nonumber \\
&-& E  \varphi_{k}''' \varphi_{-k} - {F'\over 2}  \varphi_{k}' \varphi_{-k} -
 E {1+u^2\over u f} \varphi_{k}'\varphi_{-k}'
  + 2 E {\wn^2\over u f^2} \varphi_{k}'\varphi_{-k}\Biggr]\,,
\label{boundary_term}
\end{eqnarray}
and $\wn \equiv \omega/2 r_0$.

\subsection{The solution for $h_{xy}$}

We now turn to finding the solution to the equation of motion 
(\ref{ele}).
Explicitly, Eq.~(\ref{ele}) reads
\begin{eqnarray}
&\,& u f^2 \, \phi_k '' - f (1+u^2)\,  \phi_k' + 
\wn^2 \, \phi_k = 
 - \gamma \, 37\, u^7 \, f^3\,  \phi_k^{IV}(r)
 + \gamma \, 74 \, u^6 \, f^2 \, (9 u^2 - 5)\,  
\, \phi_k '''(r) \nonumber \\
&-& \gamma \, {u^3\, f\over 8} \left(
 - 600 + 6193\, u^2 - 25552\, u^4 + 22327\, u^6 + 592\, u^3\, \wn^2 \right)\, 
\phi_k ''(r) \nonumber \\ &-& 
\gamma \, {u^2\, f\over 8} ( - 600 + 819\, u^2 - 3440\, u^4
 + 7105\, u^6 + 2368\, u^3\, \wn^2)
 \phi_k '(r) 
\nonumber \\ 
&-& \gamma \, {u^2 \over 8 f} \Biggl[ 100\, u - 550\, u^5 + 1000\, u^7
 - 450\, u^9
+ 600\, \wn^2 - 719\, u^2\, \wn^2 + 3968\, u^4\, \wn^2 \nonumber \\
&-& 2665\, u^6\, \wn^2 + 4\, u^3 ( 74 \wn^4 - 25)\Biggr]\, \phi_k(r).
\label{eqinu}
\end{eqnarray}
where we put all $\gamma$-dependent terms to the right.
Eq. (\ref{eqinu}) can be solved perturbatively in $\gamma$ by writing 
\begin{equation}
\varphi_k(u) = \varphi_k^{(0)}(u) + \gamma \varphi_k^{(1)}(u)\,,
\end{equation}
where $ \varphi_k^{(0)}(u)$ is the solution to the minimally coupled
massless scalar equation (i.e. Eq.~(\ref{eqinu}) with $\gamma=0$)
found in \cite{Policastro:2002se}, 
\begin{equation}
 \varphi_k^{(0)}(u) = (1-u)^{-{i \wn\over 2}} \left( 1 - {i \wn\over 2} \ln
{1+u\over 2} + O (\wn^2)\right)\,.
\end{equation}
The full solution obeying the incoming wave boundary condition 
at the horizon $u=1$ \cite{Policastro:2002se}
and normalized to one at the boundary $u=0$ is
\begin{equation}
 \varphi_k(u) = (1-u)^{-{i \wn\over 2}} G(u)\,,
\label{solution}
\end{equation}
where the function $G(u)$ is regular at $u=1$, and is given explicitly by
\begin{eqnarray}
G(u) &=& 1  - \gamma {25\, u^4\,  (1+u^2)\over 16} - {i\,  \wn \over 2}
\left[ 1 - {25\over 16} \gamma\,  u^4\,  (1+u^2)\right] \log{(1+u)}
\nonumber \\ &+&
 \gamma {i\, \wn\,  u^2\over 2} \left( 43\,  u^4 + 135\, u^2 + 195 \right) + O(\gamma^2, \wn^2)\,.
\end{eqnarray}

\section{ Coupling constant correction to shear viscosity}
\label{section_correction}

Having found the solution for a gravitational perturbation, 
we can compute the correlation function
 $G_{xy,xy}(\omega,q)$ by applying the  Minkowski AdS/CFT prescription
\cite{Son:2002sd}
 \begin{equation}
 G_{xy,xy}^R(\omega,q) = \lim_{u\rightarrow 0}  2 {\cal F}_k\,,
\label{prescription}
\end{equation}
where ${\cal F}_k$ is the boundary term (\ref{boundary_term}). 
 (It is assumed that all contact and momentum-independent terms 
in (\ref{prescription}) should be discarded.)
Computing the limit, to the lowest order in $\wn$ we find
 \begin{equation}
 \lim_{u\rightarrow 0}  2 {\cal F}_k =  { N^2_c r_0^4\over 4 \pi^2}
\left( 1+75 \gamma -{1\over u^2} - i \wn (1+195 \gamma) + O(\wn^2) \right)\,.
\label{limit}
\end{equation}
On a technical side, we note that the asymptotic behavior of the 
coefficients (\ref{eq:a1})--(\ref{coeffi}) and the regularity
of the solution  (\ref{solution})  at $u\rightarrow 0$ imply that 
only the first two terms in (\ref{boundary_term})
contribute to the limit  (\ref{limit}), with the 
nontrivial contribution
coming only from the term  
$\left( B - A\right) \varphi_{k}' \varphi_{-k}$.
Moreover, contributions to the coefficients $A$ and $B$ coming from 
the Weyl term $W$ vanish in the limit. The $W$ term therefore 
plays the role of a ``grey cardinal'' influencing the result indirectly
through the correction to the black brane metric and the equation of motion.

Thus the retarded correlator at zero spatial momentum and 
to the leading order in $\omega$ is
 \begin{equation}
 G_{xy,xy}^R(\omega,0) = - i \, {N^2_c r_0^3 \omega\over 8 \pi^2}
  \left( 1 + 195 \gamma \right) =  - i\,  {\pi N^2_c T^3 \omega\over 8}
  \left( 1 + 150 \gamma \right)\,,
\end{equation}
where we used the expression  (\ref{hawking}) 
for the Hawking temperature of the $\alpha'$-corrected metric.
Applying the Kubo formula (\ref{RA}), we immediately find
 \begin{equation}
\eta = {\pi\over 8} N^2_c T^3 \left( 1 + 150 \gamma\right)\,.
\end{equation}
Combining (\ref{alpha}) and (\ref{gamma}), we obtain\footnote{See footnote \ref{ftn} 
regarding the gauge coupling constant normalization.}
 \begin{equation}
\eta = {\pi\over 8} N^2_c T^3 \left( 1 + {75\over 4}\, \zeta(3)\,
(g^2_{YM} N_c)^{-3/2} +\cdots 
 \right)\,.
\label{result_eta}
\end{equation}
Using the result of Gubser, Klebanov and Tseytlin 
 (\ref{f_strong}) for the 
entropy density we find the strong coupling expansion for the ratio
 $\eta/s$
 \begin{equation}
{\eta\over s} = {1\over 4\pi} \left( 1 + 135 \gamma\right)\,.
\end{equation}
Thus for large  't Hooft coupling $g^2_{YM} N_c \gg 1$ the correction to
the ratio of shear viscosity to the entropy density in 
${\cal N}=4$ supersymmetric Yang-Mills theory is {\it positive},
 \begin{equation}
{\eta\over s} = {1\over 4\pi} \left( 1 + {135\over 8}\, \zeta(3)\, 
( g^2_{YM} N_c)^{-3/2} +\cdots
 \right)\,.
\label{result_ratio}
\end{equation}
Formulas (\ref{result_eta}) and (\ref{result_ratio}) are the main result
of this paper.  Thus we have shown that the conjecture of
\cite{Kovtun:2003wp} that $\eta/s\ge 1/4\pi$ remains valid to the next
order in the 't~Hooft coupling expansion.

\begin{acknowledgments}
A.B. would like to thank C.~Herzog and R.~Myers for 
useful discussions, and California Institute of Technology, 
Stanford University and University of Kentucky for their kind 
hospitality during the final stage of this project.
A.B. research at  Perimeter Institute is supported in part 
by funds from NSERC of Canada. A.B. is further supported by an 
NSERC Discovery grant. 
The work of J.T.L. was supported in part by DOE Grant No. DE-FG02-95ER40899.
A.O.S. would like to thank D.T.~Son and L.~G.~Yaffe 
for helpful 
discussions, and Michigan Center 
for Theoretical Physics and Perimeter Institute for their kind 
hospitality during the initial stage of this project.
  The work of A.O.S. is supported, in part, by DOE Grant No.\
DE-FG02-00ER41132. 

\end{acknowledgments}

\appendix

\section{Coefficients of the effective action}
\label{section_appendix}

The coefficients of the five-dimensional
 effective action  (\ref{s2}) are given by
\begin{eqnarray}
\label{eq:a1}
A &=& {\pi^4 T^4\over 2 u} \Biggl[ 8 f(u) + \gamma u^2 \left( -600 + 25 u^2 + 1760 u^4 - 1185 u^6 - 44 u^3 \wn^2 \right)\Biggr]\,, \\
B &=& {\pi^4 T^4\over 8 u}  \Biggl[ 24 f(u)  + \gamma u^2 \left(
-1800 +179 u^2 + 5424 u^4 - 3387 u^6 - 768 u^3 \wn^2 \right)\Biggr]\,, \\
C &=&  - {\pi^4 T^4\over 4 u^2 f} \Biggl[ 8 f(u) (3+u^2) 
+   \gamma u^2 \Biggl(  600 -825 u^2 - 16945 u^4 + 34005 u^6 - 16835 u^8 
\nonumber \\ &+& 104 u^3 \wn^2 + 872 u^5 \wn^2 \Biggr)\Biggr]\,, \\
D &=&   {\pi^4 T^4\over 8 u^3 f^2} \Biggl[ 16 f(u)^2 + 8 u f(u) \wn^2 
 +  \gamma u^3
 \Biggl( 250 u + 1570 u^5 + 2220 u^7 - 1920 u^9 \nonumber \\ &+& 600\wn^2 +25 u^2 \wn^2 - 2400 u^4 \wn^2 + 1007 u^6 \wn^2+ 8 u^3 (37 \wn^4 - 265) \Biggr) \Biggr]\,, \\
E &=&   \gamma \, 37\,  \pi^4 T^4  u^5 f(u)^2\,, \\
F &=&  \gamma\,  2 \pi^4 T^4 u^4 f(u) \, (11-37 u^2)\,. 
\label{coeffi}
\end{eqnarray}


\begin{thebibliography}{99}


\bibitem{landau}
L.D.~Landau and E.M.~Lifshitz, {\em Fluid mechanics}, Pergamon Press,
New York 1987, 2nd ed.

\bibitem{Jeon:if}
S.~Jeon,
``Hydrodynamic Transport Coefficients In Relativistic Scalar Field Theory,''
Phys.\ Rev.\ D {\bf 52}, 3591 (1995)
[arXiv:hep-ph/9409250].


\bibitem{Jeon:1995zm}
S.~Jeon and L.~G.~Yaffe,
``From Quantum Field Theory to Hydrodynamics: Transport Coefficients and
Effective Kinetic Theory,''
Phys.\ Rev.\ D {\bf 53}, 5799 (1996)
[arXiv:hep-ph/9512263].

\bibitem{Arnold:2000dr}
P.~Arnold, G.~D.~Moore and L.~G.~Yaffe,
``Transport coefficients in high temperature gauge theories. I:  Leading-log
results,''
JHEP {\bf 0011}, 001 (2000)
[arXiv:hep-ph/0010177];
``Transport coefficients in high temperature gauge theories. II: Beyond
leading log,''
JHEP {\bf 0305}, 051 (2003)
[arXiv:hep-ph/0302165].

\bibitem{Wang:2002nb}
E.~Wang and U.~W.~Heinz,
``Shear viscosity of hot scalar field theory in the real-time formalism,''
Phys.\ Rev.\ D {\bf 67}, 025022 (2003)
[arXiv:hep-th/0201116].

\bibitem{Teaney:2003pb}
D.~Teaney,
``Effect of shear viscosity on spectra, elliptic flow, and Hanbury Brown-Twiss
radii,''
Phys.\ Rev.\ C {\bf 68}, 034913 (2003).

\bibitem{Shuryak:2003xe}
E.~Shuryak,
``Why does the quark gluon plasma at RHIC behave as a nearly ideal fluid?,''
Prog.\ Part.\ Nucl.\ Phys.\  {\bf 53}, 273 (2004)
[arXiv:hep-ph/0312227].

\bibitem{Molnar:2001ux}
D.~Molnar and M.~Gyulassy,
``Saturation of elliptic flow at RHIC: Results from the covariant elastic
parton cascade model MPC,''
Nucl.\ Phys.\ A {\bf 697}, 495 (2002)
[Erratum-ibid.\ A {\bf 703}, 893 (2002)]
[arXiv:nucl-th/0104073].

\bibitem{Karsch:1986cq}
F.~Karsch and H.~W.~Wyld, 
``Thermal Green's Functions And Transport Coefficients On The Lattice,''
Phys.\ Rev.\ D {\bf 35}, 2518 (1987).


\bibitem{Aarts:2002cc}
G.~Aarts and J.~M.~Martinez Resco,
``Transport coefficients, spectral functions and the lattice,''
JHEP {\bf 0204}, 053 (2002)
[arXiv:hep-ph/0203177],
``Transport coefficients from the lattice?,''
Nucl.\ Phys.\ Proc.\ Suppl.\  {\bf 119}, 505 (2003)
[arXiv:hep-lat/0209033].

\bibitem{Gupta:2003zh}
S.~Gupta,
``The electrical conductivity and soft photon emissivity of the QCD plasma,''
Phys.\ Lett.\ B {\bf 597}, 57 (2004)
[arXiv:hep-lat/0301006].
 
 
\bibitem{Nakamura:2004sy}
A.~Nakamura and S.~Sakai,
``Transport coefficients of gluon plasma,''
arXiv:hep-lat/0406009.



\bibitem{Maldacena:1997re}
J.~M.~Maldacena,
``The large N limit of superconformal field theories and supergravity,''
Adv.\ Theor.\ Math.\ Phys.\  {\bf 2}, 231 (1998)
[Int.\ J.\ Theor.\ Phys.\  {\bf 38}, 1113 (1999)]
[arXiv:hep-th/9711200].

\bibitem{Gubser:1998bc}
S.~S.~Gubser, I.~R.~Klebanov and A.~M.~Polyakov,
``Gauge theory correlators from non-critical string theory,''
Phys.\ Lett.\ B {\bf 428}, 105 (1998)
[arXiv:hep-th/9802109].

\bibitem{Witten:1998qj}
E.~Witten,
``Anti-de Sitter space and holography,''
Adv.\ Theor.\ Math.\ Phys.\  {\bf 2}, 253 (1998)
[arXiv:hep-th/9802150].



\bibitem{Aharony:1999ti}
O.~Aharony, S.~S.~Gubser, J.~M.~Maldacena, H.~Ooguri and Y.~Oz,
``Large N field theories, string theory and gravity,''
Phys.\ Rept.\  {\bf 323}, 183 (2000)
[arXiv:hep-th/9905111].



\bibitem{Gubser:1996de}
S.~S.~Gubser, I.~R.~Klebanov and A.~W.~Peet,
``Entropy and Temperature of Black 3-Branes,''
Phys.\ Rev.\ D {\bf 54}, 3915 (1996)
[arXiv:hep-th/9602135].



\bibitem{Gubser:1998nz}
S.~S.~Gubser, I.~R.~Klebanov and A.~A.~Tseytlin,
``Coupling constant dependence in the thermodynamics of N = 4  supersymmetric
Yang-Mills theory,''
Nucl.\ Phys.\ B {\bf 534}, 202 (1998)
[arXiv:hep-th/9805156].


\bibitem{Fotopoulos:1998es}
A.~Fotopoulos and T.~R.~Taylor,
``Comment on two-loop free energy in N = 4 supersymmetric Yang-Mills  theory at
finite temperature,''
Phys.\ Rev.\ D {\bf 59}, 061701 (1999)
[arXiv:hep-th/9811224]; 
M.~A.~Vazquez-Mozo,
``A note on supersymmetric Yang-Mills thermodynamics,''
Phys.\ Rev.\ D {\bf 60}, 106010 (1999)
[arXiv:hep-th/9905030];
C.~j.~Kim and S.~J.~Rey,
``Thermodynamics of large-N super Yang-Mills theory and AdS/CFT
correspondence,''
Nucl.\ Phys.\ B {\bf 564}, 430 (2000)
[arXiv:hep-th/9905205];
A.~Nieto and M.~H.~G.~Tytgat,
``Effective field theory approach to N = 4 supersymmetric Yang-Mills at  finite
temperature,''
arXiv:hep-th/9906147.

\bibitem{Policastro:2001yc}
G.~Policastro, D.~T.~Son and A.~O.~Starinets,
``The shear viscosity of strongly coupled N = 4 supersymmetric Yang-Mills
plasma,''
Phys.\ Rev.\ Lett.\  {\bf 87}, 081601 (2001)
[arXiv:hep-th/0104066].


\bibitem{Policastro:2002se}
G.~Policastro, D.~T.~Son and A.~O.~Starinets,
``From AdS/CFT correspondence to hydrodynamics,''
JHEP {\bf 0209}, 043 (2002)
[arXiv:hep-th/0205052].

\bibitem{Policastro:2002tn}
G.~Policastro, D.~T.~Son and A.~O.~Starinets,
``From AdS/CFT correspondence to hydrodynamics. II: Sound waves,''
JHEP {\bf 0212}, 054 (2002)
[arXiv:hep-th/0210220].


\bibitem{Kovtun:2003wp}
P.~Kovtun, D.~T.~Son and A.~O.~Starinets,
``Holography and hydrodynamics: Diffusion on stretched horizons,''
JHEP {\bf 0310}, 064 (2003)
[arXiv:hep-th/0309213].

\bibitem{Herzog:2002fn}
C.~P.~Herzog,
``The hydrodynamics of M-theory,''
JHEP {\bf 0212}, 026 (2002)
[arXiv:hep-th/0210126].

\bibitem{Herzog:2003ke}
C.~P.~Herzog,
``The sound of M-theory,''
Phys.\ Rev.\ D {\bf 68}, 024013 (2003)
[arXiv:hep-th/0302086].



\bibitem{Son:2002sd}
D.~T.~Son and A.~O.~Starinets,
``Minkowski-space correlators in AdS/CFT correspondence: Recipe and
applications,''
JHEP {\bf 0209}, 042 (2002)
[arXiv:hep-th/0205051].

\bibitem{Nunez:2003eq}
A.~Nunez and A.~O.~Starinets,
``AdS/CFT correspondence, quasinormal modes, and thermal correlators in N  = 4
SYM,''
Phys.\ Rev.\ D {\bf 67}, 124013 (2003)
[arXiv:hep-th/0302026].



\bibitem{Buchel:2003tz}
A.~Buchel and J.~T.~Liu,
``Universality of the shear viscosity in supergravity,''
arXiv:hep-th/0311175.



\bibitem{Kovtun:2004de}
P.~Kovtun, D.~T.~Son and A.~O.~Starinets,
``A viscosity bound conjecture,''
arXiv:hep-th/0405231.


\bibitem{Das:1996we}
S.~R.~Das, G.~W.~Gibbons and S.~D.~Mathur,
``Universality of low energy absorption cross sections for black holes,''
Phys.\ Rev.\ Lett.\  {\bf 78}, 417 (1997)
[arXiv:hep-th/9609052].


\bibitem{Kovtun:2003vj}
P.~Kovtun and L.~G.~Yaffe,
``Hydrodynamic fluctuations, long-time tails, and supersymmetry,''
Phys.\ Rev.\ D {\bf 68}, 025007 (2003)
[arXiv:hep-th/0303010].

\bibitem{Herzog:2002pc}
C.~P.~Herzog and D.~T.~Son,
``Schwinger-Keldysh propagators from AdS/CFT correspondence,''
JHEP {\bf 0303}, 046 (2003)
[arXiv:hep-th/0212072].

\bibitem{Policastro:2001yb}
G.~Policastro and A.~Starinets,
``On the absorption by near-extremal black branes,''
Nucl.\ Phys.\ B {\bf 610}, 117 (2001)
[arXiv:hep-th/0104065].


\bibitem{kiritsis}
E.~Kiritsis, {\it String Theory in a Nutshell,} 
Princeton University Press, Princeton, 2007.



\bibitem{Grisaru:vi}
M.~T.~Grisaru and D.~Zanon,
``Sigma Model Superstring Corrections To The Einstein-Hilbert Action,''
Phys.\ Lett.\ B {\bf 177}, 347 (1986);
M.~D.~Freeman, C.~N.~Pope, M.~F.~Sohnius and K.~S.~Stelle,
``Higher Order Sigma Model Counterterms And The Effective Action For
Superstrings,''
Phys.\ Lett.\ B {\bf 178}, 199 (1986);
Q.~H.~Park and D.~Zanon,
``More On Sigma Model Beta Functions And Low-Energy Effective Actions,''
Phys.\ Rev.\ D {\bf 35}, 4038 (1987).


\bibitem{Gross:1986iv}
D.~J.~Gross and E.~Witten,
``Superstring Modifications Of Einstein's Equations,''
Nucl.\ Phys.\ B {\bf 277}, 1 (1986).

\bibitem{deHaro:2003zd}
S.~de Haro, A.~Sinkovics and K.~Skenderis,
``On alpha' corrections to D-brane solutions,''
Phys.\ Rev.\ D {\bf 68}, 066001 (2003)
[arXiv:hep-th/0302136].

\bibitem{Bremer:1998zp}
M.~S.~Bremer, M.~J.~Duff, H.~Lu, C.~N.~Pope and K.~S.~Stelle,
``Instanton cosmology and domain walls from M-theory and string theory,''
Nucl.\ Phys.\ B {\bf 543}, 321 (1999)
[arXiv:hep-th/9807051].

\bibitem{Gibbons:1976ue}
G.~W.~Gibbons and S.~W.~Hawking,
``Action Integrals And Partition Functions In Quantum Gravity,''
Phys.\ Rev.\ D {\bf 15}, 2752 (1977).



\bibitem{Pawelczyk:1998pb}
J.~Pawelczyk and S.~Theisen,
``AdS(5) x S(5) black hole metric at O(alpha'**3),''
JHEP {\bf 9809}, 010 (1998)
[arXiv:hep-th/9808126].


\end{thebibliography}
\end{document}